\documentclass[a4paper,11pt]{article}
\usepackage{indentfirst,latexsym,bm,color}
\usepackage{amsmath,amssymb,amsfonts}

\usepackage[top=1in,left=1in,right=1in]{geometry}

\makeatletter

\@addtoreset{equation}{section} \makeatother
\begin{document}
\newtheorem{dingli}{Theorem}[section]
\newtheorem{dingyi}[dingli]{Definition}
\newtheorem{tuilun}[dingli]{Corollary}
\newtheorem{zhuyi}[dingli]{Remark}
\newtheorem{yinli}[dingli]{Lemma}

\title{The Hilbert-Schmidt norms of quantum channels and matrix integrals over the unit sphere
\thanks{ This work was supported by NSF of
China (No: 11671242)}}
\author{Yuan Li,$^a$\thanks{E-mail address:
 liyuan0401@aliyun.com} \ \ \ Zhengli Chen,$^a$\thanks{E-mail address: czl@snnu.edu.cn} \ \ \  Zhihua Guo,$^a$\thanks{E-mail address: guozhihua@snnu.edu.cn} \ \ \ Yongfeng Pang$^b$\thanks{E-mail address: pangyongfengyw@xauat.edu.cn}}
\date{} \maketitle\begin{center}
\begin{minipage}{16cm}
{ \small $$     a  \  \  \ School \ of \ Mathematics \ and \
statistics,\ Shaanxi \ Normal \ University, $$ $$ Xi'an,\
710062,\ China. $$ }{\small$$ b\ \ \   \ School \ of \ Science,   \ Xi'an \ University \ of \ Architecture \ and \ Technology, $$ $$\ Xi'an, \ 710055,  \ China $$}

\end{minipage}
\end{center}
 \vspace{0.05cm}
\begin{center}
\begin{minipage}{16cm}
{\small {\bf Abstract }  The dynamics of quantum systems are generally described
by a family of quantum channels (linear, completely positive and trace preserving
maps). In this note, we mainly study the range of all possible values of $\|\mathcal{E}\|_2^2+\|\widetilde{\mathcal{E}}\|_2^2$ for quantum channels $\mathcal{E}$ and give the equivalent characterizations for quantum channels that achieve these maximum and minimum values, respectively, where $\|\mathcal{E}\|_2$ is the Hilbert-Schmidt norm of $\mathcal{E}$ and $\widetilde{\mathcal{E}}$ is a complementary channel of $\mathcal{E}.$ Also, we get a concrete description of completely positive maps on infinite dimensional systems preserving pure states. Moreover, the equivalency of several matrix integrals over the unit sphere is demonstrated and some extensions of these matrix integrals are obtained.\\}
\endabstract
\end{minipage}\vspace{0.10cm}
\begin{minipage}{16cm}

\end{minipage}
\end{center}
\begin{center} \vspace{0.01cm}
\end{center}

\section{\textbf{Introduction and Preliminary}}

It is one of the fundamental properties of quantum mechanics that the evolution of quantum states
is described by linear maps which are completely positive and trace preserving. Moreover, the combination of completely positive maps and positive maps
is a powerful tool for characterizing quantum entanglement ([15,21]). It is generally known that quantum systems are described by complex separable Hilbert spaces $\mathcal{H}$ and $\mathcal{K}.$ Let $\mathcal{B(H,K)}$
be the Banach space of all bounded linear operators from $\mathcal{H}$ into $\mathcal{K}$ with the operator norm $\|\cdot\|$ and $\mathcal{T(H)}$
be the Banach spaces of all trace class operators on $\mathcal{H}$ with the trace norm $\|\cdot\|_1.$ As usual, $\mathcal{B(H)}=\mathcal{B(H,H)},$ $A^*$ is the adjoint operator and $|A|:=(A^*A)^\frac{1}{2}$ is the absolute value of $A\in\mathcal{B(H,K)}.$ Moreover, if $x,y\in{\mathcal{H}},$ then $xy^*$ is the rank-one operator in $\mathcal{B}(\mathcal{H})$
defined by $xy^*(z)=\langle z,y\rangle x$ for vectors $z\in\mathcal{H},$ where $\langle \cdot, \cdot\rangle$ is the inner product on $\mathcal{H}.$  If $\rho\in\mathcal{T(H})$ is positive with ${\rm tr}(\rho)=1,$ then $\rho$ is called a quantum state. In particular, $\phi\phi^*$ is said to be a pure state induced by a unit vector $\phi\in\mathcal{H}.$  A quantum channel with input system  $\mathcal{H}$  and output system  $\mathcal{K}$
  is represented (in the Schr\"{o}dinger picture) by a completely
positive and trace-preserving map
$\Phi:$ ${\mathcal{T(H)}}\longrightarrow {\mathcal{T(K)}},$ which can be characterized by the operator sum representation  \begin{equation}\Phi(X)=\sum_{i=1}^{\infty}A_iXA_i^* \ \ \ \  \  \hbox { } \hbox { for all } X\in\mathcal{T(H)},\end{equation} where $A_i\in\mathcal{B(H,K)}$ with $\sum_{i=1}^{\infty}A_i^*A_i=I$ in the strong operator topology ([9,14,16]).

 Given a positive and trace preserving linear
map $\Theta:\mathcal{M}_n\longrightarrow\mathcal{M}_d$ between matrix spaces, where $\mathcal{M}_n$ denote the space of $n\times n$ matrices. Then $\Theta$ is regarded as an operator that can induce various norms,   such as $p$-norm with $1\leq p\leq\infty.$ The $p$-norm of a matrix $A\in\mathcal{M}_n$ is defined as $\|A\|_p={\rm tr}(|A|^p)^\frac{1}{p}$ for $1\leq p<\infty$ and $\|A\|_\infty$ is the operator norm $\|A\|$ of $A.$ Let $$\|\Theta\|_{p\rightarrow p}=\sup\{\frac{\|\Theta(A)\|_{p}}{\|A\|_{p}}:\  \hbox{ for all } A\neq 0\}.$$ Then $\|\Theta\|_{p\rightarrow p}$ is the norm of $\Theta,$ which is seen as an operator between Banach  spaces $\mathcal{M}_n$ and $\mathcal{M}_d$ induced by $p$-norms. In [20],  it is known that
$\|\Theta\|_{1\rightarrow 1}=1,$ $\|\Theta\|_{\infty\rightarrow \infty}=\|\Theta(I)\|_{\infty}\leq n$ and
\begin{equation}\|\Theta\|_{p\rightarrow p}\leq \|\Theta\|_{1\rightarrow 1}^{\frac{1}{p}} \|\Theta\|_{\infty\rightarrow \infty}^{1-\frac{1}{p}}\leq n^{1-\frac{1}{p} }.\end{equation} Some other interesting properties of $\|\Theta\|_{p\rightarrow p}$ have also been obtained in [13, 27]. We observe that $$\|\Theta\|_{p\rightarrow p}\geq (\frac{n}{d})^{1-\frac{1}{p}}\  \hbox{ for }\ 1\leq p\leq\infty.$$ Indeed, $\|\Theta\|_{p\rightarrow p}\geq \frac{\|\Theta(I)\|_{p}}{\|I\|_{p}}$ and Cauchy-Schwarz inequality imply $$
d^{1-\frac{1}{p}}\|\Theta(I)\|_{p}\geq {\rm tr}[\Theta(I)]={\rm tr}(I)=n,$$  so  \begin{equation}\|\Theta\|_{p\rightarrow p}\geq \frac{n}{n^\frac{1}{p}d^{1-\frac{1}{p}}}=(\frac{n}{d})^{1-\frac{1}{p}}.\end{equation}
Particularly, $$\|\Theta\|_{\infty\rightarrow \infty}\geq \frac{n}{d}   \ \ \ \hbox{ and }\ \|\Theta\|_{2\rightarrow 2}\geq \sqrt{\frac{n}{d}} .$$ Moreover, it is easy to verify that the completely depolarizing channel $\mathcal{E}_c:\mathcal{M}_n\longrightarrow\mathcal{M}_d$ with $\mathcal{E}_c(X):=\frac{{\rm tr}(X)I}{d}$ satisfies $$\|\mathcal{E}_c\|_{\infty\rightarrow \infty}=\frac{n}{d} \  \ \hbox{ and  }\ \ \|\mathcal{E}_c\|_{2\rightarrow 2}=\sqrt{\frac{n}{d}}.$$

If $\mathcal{H}$ and $\mathcal{K}$ are finite dimensional complex Hilbert spaces, we always assume that $\dim\mathcal{H}=n$ and $\dim\mathcal{K}=d.$ It is well-known that $\mathcal{B(H,K)}$ is a $dn$ dimensional Hilbert space with respect to the inner product $\langle A,B\rangle={\rm tr}(AB^*)$ for all $A,B\in\mathcal{B(H,K)}.$
Especially, $\|A\|_2=\sqrt{\langle A,A\rangle}=\sqrt{{\rm tr}(AA^*)}$ is called the Hilbert-Schmidt norm (or the Frobenius norm) of $A.$
 In this case, a quantum channel $\mathcal{E}:\mathcal{B(H)}\longrightarrow\mathcal{B(K)}$ has the Kraus decomposition $\mathcal{E}(X)=\sum\limits_{i=1}^{m} A_iX A_i^*$ with $m\leq dn.$
Clearly, $\mathcal{E}$ can be seen as an operator between Hilbert spaces $\mathcal{B(H)}$ and $\mathcal{B(K)},$ and $$\langle X,\mathcal{E}^*(Y)\rangle=\langle\mathcal{E}(X),Y\rangle={\rm tr}[\mathcal{E}(X)Y^*]={\rm tr}[\sum\limits_{i=1}^{m} A_iX A_i^*Y^*]=\langle X, \sum\limits_{i=1}^{m}A_i^*YA_i\rangle$$ for all $X\in\mathcal{B(H)}$ and $Y\in\mathcal{B(K)}.$
Thus the adjoint operator of $\mathcal{E}$ is $\mathcal{E}^*,$ which has the form $\mathcal{E}^*(Y)=\sum\limits_{i=1}^{m}A_i^*YA_i.$

The Stinespring theorem implies the existence of a Hilbert space $\mathcal{K}_1$ and an isometry $V:\mathcal{H}\rightarrow\mathcal{K}\otimes\mathcal{K}_1$ such that
\begin{equation}\mathcal{E}(X)={\rm tr}_{\mathcal{K}_1}(VXV^*)\ \ \ \ \hbox{ for all } X\in\mathcal{B(H)},\end{equation} where ${\rm tr}_{\mathcal{K}_1}(A_1\otimes A_2)={\rm tr}(A_2)A_1$ is the partial trace with respect to the second Hilbert space  ${\mathcal{K}_1}.$
The channel $\widetilde{\mathcal{E}}:\mathcal{B(H)}\rightarrow \mathcal{B(K}_1),$ which is defined by \begin{equation} \widetilde{\mathcal{E}}(X)={\rm tr}_{\mathcal{K}}(VXV^*)\ \ \ \ \hbox{ for all } X\in\mathcal{B(H)},\end{equation}
is called a complementary channel of $\mathcal{E}$ ([11,13]). Suppose that $\{e_i\}_{i=1}^m$ is an orthonormal basis of $\mathcal{K}_1.$ We conclude from equation (1.5) that the complementary channel $\widetilde{\mathcal{E}}$ can be written as
\begin{equation} \widetilde{\mathcal{E}}(X)=\sum\limits_{i,j=1}^{m}{\rm tr}(A_iXA_j^*) e_ie_j^* \ \ \ \ \hbox{ for all } X\in\mathcal{B(H)}.\end{equation} The complementary channel of a quantum channel is not unique.
Indeed, if $ \widetilde{\mathcal{E}}:\mathcal{B(H)}\rightarrow \mathcal{B(K}_1)$ is complementary to $\mathcal{E}$ and $\dim \mathcal{K}_1$ is equal to the Choi rank of the channel $\mathcal{E},$ then any other channel $\Phi:\mathcal{B(H)}\rightarrow \mathcal{B(K}_2)$ is complementary to $ \mathcal{E}$ if and only if there exists an isometry
 $V\in\mathcal{B(K}_{1},\mathcal{K}_2)$ such that $\Phi(X)=V \widetilde{\mathcal{E}}(X)V^*$ ([11,24]).

 Let $\{X_j : j=1, 2,\cdots, n^2\}$ be an orthonormal basis of $\mathcal{B(H)}.$ Then $\mathcal{E}$ is an operator between Hilbert spaces $\mathcal{B(H)}$ and $\mathcal{B(K)}$ with the Hilbert-Schmidt norm
$$\|\mathcal{E}\|_2^2={\rm tr}(\mathcal{E}^*\mathcal{E})=\sum\limits_{j=1}^{n^2}\langle\mathcal{E}(X_j),\mathcal{E}(X_j)
\rangle=\sum\limits_{j=1}^{n^2}\|\mathcal{E}(X_j)\|_2^2$$  and the complementary channel  $\Phi$ is an operator from $\mathcal{B(H)}$ into $\mathcal{B(K}_2)$ with the Hilbert-Schmidt norm
$$\|\Phi\|_2^2={\rm tr}(\Phi^*\Phi)=\|\widetilde{\mathcal{E}}\|_2^2=\sum\limits_{k=1}^{n^2}\|\widetilde{\mathcal{E}}(X_k)\|_2^2.$$

Based on inequalities (1.2) and (1.3), we want to consider the range of all possible values of the Hilbert-Schmidt norm of the quantum channel $\mathcal{E}:\mathcal{B(H)}\rightarrow \mathcal{B(K)}.$  In [17, Proposition 1], it is essentially shown that \begin{equation}\left\{
    \begin{array}{l}
      \frac{n}{d}\leq\|\mathcal{E}\|_2^2\leq n^2 \ \ \ \ \ \ \ \  \  \ \  \  \  \ \ \ \ \ \ \ \ \ \ \hbox{ if } n\leq d \ \  \ \ \ \   \\  \frac{n}{d}\leq\|\mathcal{E}\|_2^2\leq d^2n_0+d'^2\ \ \ \ \ \ \ \ \ \ \ \ \ \hbox{ if } n>d,\  \ \ \ \end{array}\right.\end{equation} and \begin{equation}\frac{n^2}{d}\leq\|\widetilde{\mathcal{E}}\|_2^2\leq n^2\end{equation} hold for all quantum channels $\mathcal{E}:\mathcal{M}_n\rightarrow \mathcal{M}_d,$  where $n_0=\lfloor \frac{n}{d}\rfloor$ (the largest integer upper bounded
by $\frac{n}{d}$) and $d'=n-n_0d.$

The aim of this note is to consider the range of all possible values of $\|\mathcal{E}\|_2^2+\|\widetilde{\mathcal{E}}\|_2^2$ for all quantum channels $\mathcal{E}:\mathcal{B(H)}\rightarrow \mathcal{B(K)}.$ Using the above two equations, we can present an upper bound and
a lower bound of $\|\mathcal{E}\|_2^2+\|\widetilde{\mathcal{E}}\|_2^2.$ However, we find that this upper bound is not optimal.  More specifically, we mainly show the following results.

{\bf Theorem 1.} Let $\mathcal{E}:\mathcal{B(H)}\longrightarrow\mathcal{B(K)}$ be a quantum channel and $\widetilde{\mathcal{E}}$ be a complementary channel of $\mathcal{E}.$ Suppose that $\dim\mathcal{H}=n$ and $\dim\mathcal{K}=d.$  Then 

$(1)$ \  $\frac{n+n^2}{d}\leq\|\mathcal{E}\|_2^2+\|\widetilde{\mathcal{E}}\|_2^2\leq n^2+n.$

$(2)$  The following statements are equivalent:  

\ \ \ \ $(a)$ \ $\|\mathcal{E}\|_2^2+\|\widetilde{\mathcal{E}}\|_2^2=\frac{n+n^2}{d};$

\ \ \ \ $(b)$ \ $\mathcal{E}$ is the completely depolarizing channel $\mathcal{E}(X)=\frac{{\rm tr}(X)I}{d};$

\ \ \ \ $(c)$ \ $\|\mathcal{E}\|_2+\|\widetilde{\mathcal{E}}\|_2=\frac{\sqrt{n}+n}{\sqrt{d}};$

\ \ \ \ $(d)$ \ $\mathcal{E}$ has the Hilbert-Schmidt norm $\|\mathcal{E}\|_2=\sqrt{\frac{n}{d}}.$

$(3)$  $\|\mathcal{E}\|_2^2+\|\widetilde{\mathcal{E}}\|_2^2=n^2+n$ if and only if
$\mathcal{E}(X)=VXV^*$  or  $\mathcal{E}(X)={\rm tr}(X)\psi\psi^*,$ where $V\in\mathcal{B(H,K)}$ is an isometry and $\psi\in\mathcal{K}$ is a unit vector.

To prove Theorem 1, we need to characterize the structures of completely positive maps on finite dimensional quantum systems preserving pure states. A similar result has been obtained in [1,7]. More generally, we give a concrete description of completely positive maps on infinite dimensional systems preserving all pure states.

{\bf Theorem 2.} Let $\Phi:\mathcal{T(H)}\longrightarrow\mathcal{T(K)}$ be a completely positive maps. Then $\Phi(xx^*)$ are pure states for all unit vectors $x\in\mathcal{H}$ if and only if $\Phi(X)=VXV^*$ or   $\Phi(X)={\rm tr}(X)\psi\psi^*,$ where $V\in\mathcal{B(H,K)}$ is an isometry and $\psi\in\mathcal{K}$ is a unit vector.

Moreover, we also have the aid of the tools of some matrices integrals over the unit sphere.
Although these formulae have been used in references [2,6,25,28], we don't find the relevant proofs. Particularly, we give the equivalent relation between the following integrals.

{\bf Proposition 3.} Let $\dim\mathcal{H}=n,$ $\mathcal{S}\subseteq\mathcal{H}$ be the unit sphere and $d\phi$ be the normalized measure on $\mathcal{S}$ induced
by the Haar measure of the unitary group.  Then the following statements are equivalent:

$(a)$  $\int_{\mathcal{S}} \phi\phi^*\otimes\phi\phi^*d\phi=\frac{I\otimes I+S}{n(n+1)},$ where $S\in\mathcal{B(H\otimes H)}$ is the swap operator, i.e., $S(x\otimes y)=y\otimes x$ for all vectors $x,y\in\mathcal{H};$

$(b)$  $\int_{\mathcal{S}}\langle A\phi,\phi\rangle\langle B\phi,\phi\rangle d\phi=\frac{{\rm tr}(AB)+{\rm tr}(A){\rm tr}(B)}{n(n+1)}$ for all $A,B\in\mathcal{B(H)};$

$(c)$ $\int_{\mathcal{S}}\langle e_{i}e_j^*\phi,\phi\rangle\phi\phi^* d\phi=\frac{\delta_{ij}I+e_{i}e_j^* }{n(n+1)},$ where $\{e_i\}_{i=1}^n$ is an orthonormal basis of $\mathcal{H}$  and $\delta_{ij}$ is Kronecker  delta;

$(d)$  $\int_{\mathcal{S}} {\rm tr}(A\phi\phi^*)\phi\phi^*d\phi=\frac{A+{\rm tr}(A)I}{n(n+1)}$ for all $A\in\mathcal{B(H)}.$

\section{Matrix integrals over the unit sphere}

When $\mathcal{H}$ is finite dimensional, the following result is known in [23] by using Schur Lemma ([5]).
However, this method is not applicable for the infinite-dimensional setting. 

{\bf Lemma 4.} Let $\Phi:\mathcal{T(H)}\rightarrow \mathcal{B(H)}$ be a linear map. If $\Phi$ is continuous with respect to the topologies $(\mathcal{T(H)},\|\cdot\|_1)$ and $(\mathcal{B(H)},\|\cdot\|),$ then $\Phi$ has the covariance property $\Phi(UXU^*)=U\Phi(X)U^*$ for all $X\in\mathcal{T(H)}$ and all unitary operators $U\in\mathcal{B(H)}$ if and only if there exist complex numbers $\lambda$ and $\mu$ such that $\Phi(X)=\lambda X+\mu {\rm tr}(X)I.$

{\bf Proof.} Sufficiency is clear.

Necessity.
Let $e\in \mathcal{H}$ be a fixed unit vector and $\Im$ be an abelian C*-algebra generated by $ee^*$ and $I.$ For all unitary operators $U\in\Im,$ we have
$$\Phi(ee^*)=\Phi(Uee^*U^*)=U\Phi(ee^*)U^*.$$ Then $\Phi(ee^*)U=U\Phi(ee^*),$ which implies $\Phi(ee^*)\in\Im,$ so $\Phi(ee^*)=\lambda ee^*+\mu I,$ where $\mu,\lambda\in \mathbb{C}.$ For any unit vector $f\in \mathcal{H},$ there exists a unitary operator $U_0$ such that $U_0e=f,$ which implies that $$\Phi(ff^*)=\Phi(U_0ee^*U_0^*)=U_0\Phi(ee^*)U_0^*=U_0(\lambda ee^*+\mu I)U_0^*=\lambda ff^*+\mu I.$$

For any positive operator $A\in\mathcal{T(H)},$ we get from the spectral decomposition theorem that $A=\sum_{j=1}^{\infty}\lambda_jx_jx_j^*,$ where $\{x_j\}_{j=1}^\infty$ is
an orthonormal basis of ${\mathcal{H}}$ and $\lambda_j\geq\lambda_{j+1}\geq0$ are eigenvalues of $A.$ Clearly, $$\lim\limits_{n\rightarrow\infty}\|A_n-A\|_1=\lim\limits_{n\rightarrow\infty}\sum_{j=n+1}^{\infty}\lambda_j=0,$$ where $A_n=\sum_{j=1}^{n}\lambda_jx_jx_j^*.$ Then
$$\begin{array}{rcl}\Phi(A)=\lim\limits_{n\rightarrow\infty}\Phi(A_n)&=&\lim\limits_{n\rightarrow\infty}\sum_{j=1}^{n}
\lambda_j\Phi(x_jx_j^*)\\&=&\lim\limits_{n\rightarrow\infty}\sum_{j=1}^{n}\lambda_j(\lambda x_jx_j^*+\mu I)\\&=&\lambda A+\mu {\rm tr}(A)I\end{array}$$ follows from equations $$\lim\limits_{n\rightarrow\infty}\|A_n-A\|\leq\lim\limits_{n\rightarrow\infty}\|A_n-A\|_1=0$$  and $$ \lim\limits_{n\rightarrow\infty}\|\sum_{j=1}^{n}\lambda_jI- {\rm tr}(A)I\|=\lim\limits_{n\rightarrow\infty}\sum_{j=n+1}^{\infty}\lambda_j=0.$$  If $A\in\mathcal{T(H)}$ is a self-adjoint operator, then $A=A^+-A^-,$ where $A^+=\frac{|A|+A}{2}$ and $A^-=\frac{|A|-A}{2}$ are positive operators. So $$\Phi(A)=\Phi(A^+)-\Phi(A^-)=\lambda A^++\mu {\rm tr}(A^+)I-(\lambda A^-+\mu {\rm tr}(A^-)I)=\lambda A+\mu {\rm tr}(A)I.$$

For any $X\in\mathcal{T}\mathcal{(H)},$ we have
$X=Re(X)+\sqrt{-1}Im(X),$ where $Re(X)=\frac{X+X^*}{2}$ and $Im(X)=\frac{X-X^*}{2\sqrt{-1}}$ are self-adjoint operators. Then $$\begin{array}{rcl}\Phi(X)&=&\Phi(Re(X)+\sqrt{-1}Im(X))\\&=&\lambda Re(X)+\mu {\rm tr}(Re(X))I+\sqrt{-1}(\lambda Im(X)+\mu {\rm tr}(Im(X))I)\\&=&\lambda X+\mu {\rm tr}(X)I.\end{array}$$
$\Box$

{\bf Remark 1.} If $\mathcal{H}$ is finite dimensional, then $\mathcal{T(H)}=\mathcal{B(H)}.$ Thus the linear map $\Phi:\mathcal{T(H)}\rightarrow \mathcal{B(H)}$ is continuous with respect to the topologies $(\mathcal{T(H)},\|\cdot\|_1)$ and $(\mathcal{B(H)},\|\cdot\|),$ so Lemma 4 is an extension of [3, Theorem 3.1] obtained by Bhat. It is  worth noting that our proof techniques are completely different.

The following corollary is a useful formula  for calculating fidelity and disturbance of quantum channels in [25]. Based on Lemma 4 and Riesz representation theorem, we shall give a new and simpler proof.

{\bf Lemma 5.} Let $A,B\in\mathcal{B(H)}$ and  $\mathcal{S}\subseteq\mathcal{H}$ be the unit sphere.
If $\dim\mathcal{H}=n$ and $d\phi$ is the normalized measure on $\mathcal{S}$ induced by the Haar measure of the unitary group, then $$\int_{\mathcal{S}}\langle A\phi,\phi\rangle\langle B\phi,\phi\rangle d\phi=\frac{{\rm tr}(AB)+{\rm tr}(A){\rm tr}(B)}{n(n+1)}.$$

{\bf Proof.}  It is well-known that $\mathcal{B(H)}$ is a Hilbert space with
the inner product $\langle C,D\rangle={\rm tr}(CD^*)$ for all $C,D\in\mathcal{B(H)}.$
Define a sesquilinear $f: \mathcal{B(H)}\times \mathcal{B(H)}\longrightarrow  \mathbb{C}$ by
$$f(C,D)=\int_{\mathcal{S}}\langle C\phi,\phi\rangle\langle D^*\phi,\phi\rangle d\phi.$$ It is easy to see that   $$|f(C,D)|\leq\int_{\mathcal{S}}|\langle C\phi,\phi\rangle\langle D^*\phi,\phi\rangle| d\phi\leq\int_{\mathcal{S}}\|C\| \|D^*\| d\phi=\|C\|\|D^*\|.$$ Using Riesz representation theorem ([10, Theorem 1.67]), we get that there is a unique linear operator $\Phi:\mathcal{B(H)}\rightarrow \mathcal{B(H)}$ such that
$$f(C,D)=\langle\Phi(C),D\rangle={\rm tr}[\Phi(C)D^*].$$
Clearly, for all unitary operators $U\in\mathcal{B(H)},$ there are $$f(U^*CU,U^*DU)=\int_{\mathcal{S}}\langle CU\phi,U\phi\rangle\langle
D^*U\phi,U\phi\rangle d\phi=\int_{\mathcal{S}}\langle C\psi,\psi\rangle\langle D^*\psi,\psi\rangle d(U^*\psi)=f(C,D),$$ so $$\langle\Phi(C),D\rangle=\langle\Phi(U^*CU),U^*DU\rangle={\rm tr}[U\Phi(U^*CU)U^*D^*]
=\langle U\Phi(U^*CU)U^*,D\rangle$$ for all unitary operators $U\in\mathcal{B(H)}.$ Thus $\Phi(C)= U\Phi(U^*DU)U^*$ for all $C\in\mathcal{B(H)}$ and unitary operators $U\in\mathcal{B(H)}.$  Then Lemma 4 implies that there exist complex numbers $\lambda$ and $\mu$ such that $\Phi(C)=\lambda C+\mu{\rm tr}(C)I.$ Therefore, \begin{equation}\int_{\mathcal{S}}\langle C\phi,\phi\rangle\langle D\phi,\phi\rangle d\phi=f(C,D^*)={\rm tr}[\Phi(C)D]=\lambda {\rm tr}(CD)+\mu {\rm tr}(C){\rm tr}(D).\end{equation}
Obviously, \begin{equation}1=\int_{\mathcal{S}}\langle  \phi,\phi\rangle\langle  \phi,\phi\rangle d\phi=f(I,I)=n\lambda +n^2\mu .\end{equation} Let $e_1$ and $e_2$ be unit vectors in $\mathcal{H}$ with $e_1\perp e_2.$
A direct calculation yields $$f(e_1e_2^*, e_1e_2^*)=\int_{\mathcal{S}}\langle e_1e_2^*\phi,\phi\rangle\langle e_2e_1^*\phi,\phi\rangle d\phi=\int_{\mathcal{S}}|\langle e_1,\phi\rangle\langle e_2,\phi\rangle|^2 d\phi=f( e_1e_1^*,e_2e_2^*).$$ Then equation (2.1) implies that
  \begin{equation}\lambda=f(e_1e_2^*, e_1e_2^*)=f( e_1e_1^*,e_2e_2^*)=\mu.\end{equation}
  Combining equations (2.2) and (2.3), we have $$\lambda=\mu=\frac{1}{n+n^2}.$$ Using equation (2.1) again, we get the desired conclusion. $\Box$

{\bf Proof of Proposition 3.} $(d)\Longrightarrow(c)$ is clear. For all $A,B\in\mathcal{B(H)},$  we know that
 \begin{equation}\langle A\otimes B,\int_{\mathcal{S}} \phi\phi^*\otimes\phi\phi^*d\phi\rangle
=\int_{\mathcal{S}}{\rm tr}[(A\otimes B)(\phi\phi^*\otimes\phi\phi^*)]d\phi=\int_{\mathcal{S}}\langle A\phi,\phi\rangle\langle B\phi,\phi\rangle d\phi \end{equation} and
 \begin{equation}\langle A\otimes B,\frac{I\otimes I+S}{n(n+1)}\rangle
=\frac{{\rm tr}(A\otimes B)+{\rm tr}[(A\otimes B)S]}{n(n+1)}=\frac{{\rm tr}(A){\rm tr}(B)+{\rm tr}(AB)}{n(n+1)},\end{equation} where $S\in\mathcal{B(H\otimes H)}$ is the swap operator.
So  $$\int_{\mathcal{S}} \phi\phi^*\otimes\phi\phi^*d\phi=\frac{I\otimes I+S}{n(n+1)}\Longleftrightarrow \int_{\mathcal{S}}\langle A\phi,\phi\rangle\langle B\phi,\phi\rangle d\phi=\frac{{\rm tr}(A){\rm tr}(B)+{\rm tr}(AB)}{n(n+1)}$$ for all $A,B\in\mathcal{B(H)}.$ That is, $(a)\Longleftrightarrow(b).$

Furthermore, for all $A,B\in\mathcal{B(H)},$ there are
$$\langle\int_{\mathcal{S}} {\rm tr}(A\phi\phi^*)\phi\phi^*d\phi,B^*\rangle=\int_{\mathcal{S}}{\rm tr}[(A\phi\phi^*) {\rm tr}(B\phi\phi^*)]d\phi=\int_{\mathcal{S}}\langle A\phi,\phi\rangle\langle B\phi,\phi\rangle d\phi$$
and $$\langle\frac{A+{\rm tr}(A)I}{n(n+1)},B^*\rangle=\frac{{\rm tr}(AB)+{\rm tr}(A){\rm tr}(B)}{n(n+1)}.$$
Thus $(b)\Longleftrightarrow(d).$

$(c)\Longrightarrow(d).$  Let $A\in\mathcal{B(H)}$ and $\{e_i\}_{i=1}^n$ be an orthonormal basis of $\mathcal{H}.$ It is obvious that $A=\sum\limits_{i,j=1}^{n}{\rm tr}(Ae_{i}e_j^*)e_{j}e_i^*,$ so
$${\rm tr}(A\phi\phi^*)\phi\phi^*=\sum\limits_{i,j=1}^{n}{\rm tr}(Ae_{i}e_j^*){\rm tr}(e_{j}e_i^*\phi\phi^*)\phi\phi^*,$$
which implies $$\begin{array}{rcl}\int_{\mathcal{S}} {\rm tr}(A\phi\phi^*)\phi\phi^*d\phi&=&\int_{\mathcal{S}}\sum\limits_{i,j=1}^{n}{\rm tr}(Ae_{i}e_j^*) {\rm tr}(e_{j}e_i^*\phi\phi^*)\phi\phi^*d\phi\\&=&\sum\limits_{i,j=1}^{n}{\rm tr}(Ae_{i}e_j^*)\int_{\mathcal{S}} {\rm tr}(e_{j}e_i^*\phi\phi^*)\phi\phi^*d\phi\\&=&\sum\limits_{i,j=1}^{n}{\rm tr}(Ae_{i}e_j^*)\frac{\delta_{ij}I+e_{j}e_i^* }{n(n+1)}\\&=&\frac{A+{\rm tr}(A)I}{n(n+1)}.\end{array}$$    $\Box$

Using Proposition 3 and Lemma 4, the following corollary is immediate. In the remaining part of this note, we always assume that $\mathcal{S}\subseteq\mathcal{H}$ be the unit sphere and $d\phi$ be the normalized measure on $\mathcal{S}$ induced
by the Haar measure of the unitary group. 

{\bf Corollary 6.}  If $\dim\mathcal{H}=n$ and $A\in\mathcal{B(H)},$  then

$(a)$  $\int_{\mathcal{S}} |{\rm tr}(A\phi\phi^*)|^2d\phi=\frac{{\rm tr}(AA^*)+|{\rm tr}(A)|^2}{n(n+1)}.$

$(b)$  $\int_{\mathcal{S}} {\rm tr}(A\phi\phi^*)d\phi=\frac{{\rm tr}(A)}{n}.$

As a generalization of the above result, we consider two integral formulas for $A\in\mathcal{B(H\otimes H)}.$
Recall that the partial trace ${\rm tr}_1(A)\in\mathcal{B(H)}$ is define by $$\langle {\rm tr}_1(A)x,y\rangle=\sum\limits_{i=1}^{n}\langle A(e_i\otimes x),e_i\otimes y\rangle$$ for all $x,y\in\mathcal{H},$ where $\{e_i\}_{i=1}^n$ is an orthonormal basis of $\mathcal{H}.$ The definition of ${\rm tr}_2(A)\in\mathcal{B(H)}$ is similar.

{\bf Corollary 7.} If $\dim\mathcal{H}=n$ and $A\in\mathcal{B(H\otimes H)},$  then

 $(a)$  $$\int_{\mathcal{S}}(\phi\phi^*\otimes I)A(\phi\phi^*\otimes I)d\phi=\frac{A+I\otimes {\rm tr}_1(A)}{n(n+1)}.$$

$(b)$  $$\int_{\mathcal{S}}\int_{\mathcal{S}} (\phi\phi^*\otimes\psi\psi^*)A(\phi\phi^*\otimes\psi\psi^*) d\phi d\psi= \frac{A+I\otimes {\rm tr}_1(A)+ {\rm tr}_2(A)\otimes I+{\rm tr}(A)I\otimes I}{n^2(n+1)^2}.$$

{\bf Proof.} Since $A\in\mathcal{B(H\otimes H)},$ we can write that $A=\sum\limits_{j=1}^{m}B_j\otimes C_j,$ where
$B_j,C_j\in\mathcal{B(H)}.$ Then ${\rm tr}_1(A)=\sum\limits_{j=1}^{m}{\rm tr}(B_j)C_j$ and ${\rm tr}_2(A)=\sum\limits_{j=1}^{m}{\rm tr}(C_j)B_j.$

 $(a)$  $$\begin{array}{rcl}\int_{\mathcal{S}}(\phi\phi^*\otimes I)A(\phi\phi^*\otimes I)d\phi&=&\int_{\mathcal{S}}(\phi\phi^*\otimes I)(\sum\limits_{j=1}^{m}B_j\otimes C_j)(\phi\phi^*\otimes I)d\phi\\&=&\sum\limits_{j=1}^{m}\int_{\mathcal{S}}(\langle B_j\phi,\phi\rangle\phi\phi^*\otimes C_j)d\phi\\&=&\sum\limits_{j=1}^{m}\int_{\mathcal{S}}\langle B_j\phi,\phi\rangle\phi\phi^*d\phi\otimes C_j\\&=&\sum\limits_{j=1}^{m}\frac{B_j+{\rm tr}(B_j)I}{n(n+1)}\otimes C_j\\&=&\frac{A+I\otimes {\rm tr}_1(A)}{n(n+1)}.\end{array}$$

(b)  $$\begin{array}{rcl}\int_{\mathcal{S}}\int_{\mathcal{S}} (\phi\phi^*\otimes\psi\psi^*)A(\phi\phi^*\otimes\psi\psi^*) d\phi d\psi&=&\int_{\mathcal{S}} \int_{\mathcal{S}}(\phi\phi^*\otimes \psi\psi^*)(\sum\limits_{j=1}^{m}B_j\otimes C_j)(\phi\phi^*\otimes \psi\psi^*)d\phi d\psi\\&=&\sum\limits_{j=1}^{m}\int_{\mathcal{S}}\int_{\mathcal{S}}(\langle B_j\phi,\phi\rangle\phi\phi^*\otimes \langle C_j\psi,\psi\rangle\psi\psi^*)d\phi d\psi\\&=&\sum\limits_{j=1}^{m}[\int_{\mathcal{S}}\langle B_j\phi,\phi\rangle\phi\phi^*d\phi\otimes \int_{\mathcal{S}}\langle C_j\psi,\psi\rangle\psi\psi^* d\psi]\\&=&\sum\limits_{j=1}^{m}[\frac{B_j+{\rm tr}(B_j)I}{n(n+1)}\otimes\frac{C_j+{\rm tr}(C_j)I}{n(n+1)}]\\&=&\frac{A+I\otimes {\rm tr}_1(A)+ {\rm tr}_2(A)\otimes I+{\rm tr}(A)I\otimes I}{n^2(n+1)^2}.\end{array}$$     $\Box$

\section{Applications of Schur-Weyl duality}

In order to obtain some more general integral formulae, we need the tools of Schur-Weyl duality ([18,29]).

 Consider the following representation of the unitary group $\mathcal{U(H)}$ on $\mathcal{H}.$ For any $U\in\mathcal{U(H)},$ we define $\Delta(U)\in \mathcal{B}(\mathcal{H}^{\otimes m})$ by
$$\Delta(U)(x_1\otimes x_2\otimes\cdots\otimes x_m)=Ux_1\otimes Ux_2\otimes\cdots\otimes U x_m$$  for all $x_1\otimes x_2\otimes\cdots\otimes x_m\in\mathcal{H}^{\otimes m}.$
 Let $\mathbb{S}_m$
be the symmetric group of degree $m.$ Consider the canonical representation of the symmetric group $\mathbb{S}_m$ on $\mathcal{H}^{\otimes m}.$
That is, $$\Gamma(s)(x_1\otimes x_2\otimes\cdots\otimes x_m)=x_{s^{-1}(1)}\otimes x_{s^{-1}(2)}\otimes\cdots\otimes x_{s^{-1}(m)},$$ for $s\in\mathbb{S}_m.$ It is clear that all $\Gamma(s)$ are unitary operators. We denote the groups $$\Delta(\mathcal{U(H)})=\{U^{\otimes m}: \ V \in \mathcal{U(H)}\} \ \hbox{ and }\ \Gamma(\mathbb{S}_m)=\{\Gamma(s): \ s \in \mathbb{S}_m\}.$$ Then Schur-Weyl duality states as follows.

{\bf Theorem (Schur-Weyl duality).}   The following two algebras are commutants of one another in  $\mathcal{B}(\mathcal{H}^{\otimes m}).$

$(a)$ $Alg\{\Delta(\mathcal{U(H)})\},$ the complex algebra spanned by $\Delta(\mathcal{U(H)}).$

$(b)$ $Alg\{\Gamma(\mathbb{S}_m)\},$ the complex algebra spanned by $\Gamma(\mathbb{S}_m).$

In [6,28], the following formulas has been used. However, there is no specific proof provided in [6]. For the convenience of readers, we give the detailed proof.

{\bf Proposition 8.} Let $\dim\mathcal{H}=n.$  Then

$$\int_{\mathcal{S}}(\phi\phi^*)^{\otimes k}d\phi
 =\frac{\sum_{s\in\mathbb{S}_k}\Gamma(s)}{n(n+1)\cdots(n+k-1)}
 =\frac{k!P_k}{n(n+1)\cdots(n+k-1)}.$$ Here $P_k=\frac{\sum_{s\in\mathbb{S}_k}\Gamma(s)}{k!}\in B(\mathcal{H}^{\otimes k})$ is an orthogonal projection and $rank (P_k)=\frac{n(n+1)\cdots(n+k-1)}{k!}.$

{\bf Proof.}  If $k=1,$ then $\int_{\mathcal{S}}\phi\phi^*d\phi
 =\frac{I}{n}$ follows from Corollary 6. For $k=2,$ we get from Proposition 3 that the desired conclusion holds.  In the following, we give the proof of $k=3.$ The proof technique of $k>3$ is analogous. For simplicity, we denote by $$\Gamma(s_1)=I^{\otimes 3},\ \ \ \Gamma(s_2)=I\otimes S \ \ \ \hbox{ and } \ \ \
\Gamma(s_3)=S\otimes I,$$  where $S\in\mathcal{B(H\otimes H)}$ is the swap operator.
Moreover, define $$\Gamma(s_4)(x_1\otimes x_2\otimes x_3)=x_3\otimes x_2\otimes x_1, \ \ \ \
\Gamma(s_5)(x_1\otimes x_2\otimes x_3)=x_3\otimes x_1\otimes x_2$$ and $$\Gamma(s_6)(x_1\otimes x_2\otimes x_3)=x_2\otimes x_3\otimes x_1$$ for all $x_1\otimes x_2\otimes x_3\in\mathcal{H}^{\otimes 3}.$
Then for any $U\in\mathcal{U(H)},$ it is clear that $$U^{\otimes 3}\int_{\mathcal{S}}\phi\phi^*\otimes\phi\phi^*\otimes\phi\phi^*d\phi (U^{\otimes 3})^*=\int_{\mathcal{S}}\phi\phi^*\otimes\phi\phi^*\otimes\phi\phi^*d\phi.$$ So Schur-Weyl duality implies that  $$\int_{\mathcal{S}}\phi\phi^*\otimes\phi\phi^*\otimes\phi\phi^*d\phi\in Alg\{\Gamma(\mathbb{S}_3)\}
=\left\{\sum_{i=1}^6\lambda_i\Gamma(s_i):\ \lambda_i\in\mathbb{C}\hbox{ for }i=1,2,\cdots,6 \right\}.$$
Thus there exist complex numbers $\lambda_1,\lambda_2,\cdots,\lambda_6$ such that \begin{equation}\int_{\mathcal{S}}\phi\phi^*\otimes\phi\phi^*\otimes\phi\phi^*d\phi=
\sum_{i=1}^6\lambda_i\Gamma(s_i).\end{equation} Therefore, \begin{equation}\begin{array}{rcl}\int_{\mathcal{S}}\phi\phi^*\otimes\phi\phi^*\otimes\phi\phi^*d\phi&=&
(I\otimes S)\int_{\mathcal{S}}\phi\phi^*\otimes\phi\phi^*\otimes\phi\phi^*d\phi\\&=&
\sum_{i=1}^6\lambda_i\Gamma(s_2)\Gamma(s_i)\\&=&\lambda_2\Gamma(s_1)+\lambda_1\Gamma(s_2)+
\lambda_6\Gamma(s_3)+\lambda_5\Gamma(s_4)+\lambda_4\Gamma(s_5)+\lambda_3\Gamma(s_4)\end{array} \end{equation}
and
\begin{equation}\begin{array}{rcl}\int_{\mathcal{S}}\phi\phi^*\otimes\phi\phi^*\otimes\phi\phi^*d\phi&=&
(S\otimes I)\int_{\mathcal{S}}\phi\phi^*\otimes\phi\phi^*\otimes\phi\phi^*d\phi\\&=&
\sum_{i=1}^6\lambda_i\Gamma(s_3)\Gamma(s_i)\\&=&\lambda_3\Gamma(s_1)+\lambda_5\Gamma(s_2)+
\lambda_1\Gamma(s_3)+\lambda_6\Gamma(s_4)+\lambda_2\Gamma(s_5)+\lambda_4\Gamma(s_4).\end{array} \end{equation}

In a similar way, we get that \begin{equation}\begin{array}{rcl}\int_{\mathcal{S}}\phi\phi^*\otimes\phi\phi^*\otimes\phi\phi^*d\phi&=&
\Gamma(s_4)\int_{\mathcal{S}}\phi\phi^*\otimes\phi\phi^*\otimes\phi\phi^*d\phi\\&=&
\sum_{i=1}^6\lambda_i\Gamma(s_4)\Gamma(s_i)\\&=&\lambda_4\Gamma(s_1)+\lambda_6\Gamma(s_2)+
\lambda_5\Gamma(s_3)+\lambda_1\Gamma(s_4)+\lambda_3\Gamma(s_5)+\lambda_2\Gamma(s_4),\end{array} \end{equation}

\begin{equation}\begin{array}{rcl}\int_{\mathcal{S}}\phi\phi^*\otimes\phi\phi^*\otimes\phi\phi^*d\phi&=&
\Gamma(s_5)\int_{\mathcal{S}}\phi\phi^*\otimes\phi\phi^*\otimes\phi\phi^*d\phi\\&=&
\sum_{i=1}^6\lambda_i\Gamma(s_5)\Gamma(s_i)\\&=&\lambda_6\Gamma(s_1)+\lambda_4\Gamma(s_2)+
\lambda_2\Gamma(s_3)+\lambda_3\Gamma(s_4)+\lambda_1\Gamma(s_5)+\lambda_5\Gamma(s_4),\end{array}\end{equation}
and
\begin{equation}\begin{array}{rcl}\int_{\mathcal{S}}\phi\phi^*\otimes\phi\phi^*\otimes\phi\phi^*d\phi&=&
\Gamma(s_6)\int_{\mathcal{S}}\phi\phi^*\otimes\phi\phi^*\otimes\phi\phi^*d\phi\\&=&
\sum_{i=1}^6\lambda_i\Gamma(s_6)\Gamma(s_i)\\&=&\lambda_5\Gamma(s_1)+\lambda_3\Gamma(s_2)+
\lambda_4\Gamma(s_3)+\lambda_2\Gamma(s_4)+\lambda_6\Gamma(s_5)+\lambda_1\Gamma(s_4).\end{array} \end{equation}
Combining equations (3.1)-(3.6), we conclude that \begin{equation} \int_{\mathcal{S}}\phi\phi^*\otimes\phi\phi^*\otimes\phi\phi^*d\phi=\frac{\sum_{i=1}^6\lambda_i}{6}
\sum_{i=1}^6\Gamma(s_i).\end{equation} It can be readily verified that $${\rm tr}[\Gamma(s_1)]=n^{3}, \ \ {\rm tr}[\Gamma(s_2)]={\rm tr}[\Gamma(s_3)]={\rm tr}[\Gamma(s_4)]=n^2  \
\hbox{ and } \ \ {\rm tr}[\Gamma(s_5)]={\rm tr}[\Gamma(s_6)]=n.$$ Thus equation (3.7) implies \begin{equation} 1={\rm tr}[\int_{\mathcal{S}}\phi\phi^*\otimes\phi\phi^*\otimes\phi\phi^*d\phi]=\frac{\sum_{i=1}^6\lambda_i}{6}
\sum_{i=1}^6{\rm tr}[\Gamma(s_i)]=\frac{\sum_{i=1}^6\lambda_i}{6}(n^3+3n^2+2n),\end{equation} which yields $$\frac{\sum_{i=1}^6\lambda_i}{6}=\frac{1}{n^3+3n^2+2n}=\frac{1}{n(n+1)(n+2)}.$$
So \begin{equation}\int_{\mathcal{S}}\phi\phi^*\otimes\phi\phi^*\otimes\phi\phi^*d\phi
 =\frac{\sum_{s\in\mathbb{S}_3}\Gamma(s)}{(n+2)(n+1)n}.\end{equation} Then $\sum_{s\in\mathbb{S}_3}\Gamma(s)$ is a  positive operator.
 It is easy to check that $$(\sum_{s\in\mathbb{S}_3}\Gamma(s))^2=6\sum_{s\in\mathbb{S}_3}\Gamma(s).$$ Therefore, $\frac{ \sum_{s\in\mathbb{S}_3}\Gamma(s)}{6}$ is an orthogonal projection with $$rank(\frac{ \sum_{s\in\mathbb{S}_3}\Gamma(s)}{6})={\rm tr}(\frac{ \sum_{s\in\mathbb{S}_3}\Gamma(s)}{6})=\frac{(n+2)(n+1)n}{6}.$$
  $\Box$

{\bf Remark 2.} In general, $\Gamma(s)$ are unitary operators for $s\in\mathbb{S}_k$ and
$\Gamma(s)$ may not be self-adjoint.
For example, $\Gamma(s_5)=\Gamma(s_6)^*$ in the proof of Proposition 8.
However, $P_k=\frac{\sum_{s\in\mathbb{S}_k}\Gamma(s)}{k!}$ is an orthogonal projection onto the symmetric subspace $(\mathcal{H}^{\otimes k})_+,$ which is defined by $(\mathcal{H}^{\otimes k})_+=\{x\in\mathcal{H}^{\otimes k}: \Gamma(s)x=x \hbox{ for all } s\in\mathbb{S}_k\}$ ([6, 26]).

 As an extension of Corollary 7,  we get the following integral formulae.

{\bf Theorem 9.}   If $\dim\mathcal{H}=n$ and $A,B\in\mathcal{B(H)},$ then

 $(a)$ $$\int_{\mathcal{S}}\langle A\phi,\phi\rangle\langle B\phi,\phi\rangle\phi\phi^*d\phi
 =\frac{[{\rm tr}(A){\rm tr}(B)+{\rm tr}(AB)]I+{\rm tr}(A)B+{\rm tr}(B)A+AB+BA}{(n+2)(n+1)n}.$$

 $(b)$  $$\int_{\mathcal{S}}\langle A\phi,\phi\rangle\phi\phi^*\otimes\phi\phi^*d\phi
 =\frac{{\rm tr}(A)(I\otimes I+S)+I\otimes A+A\otimes I+S(A\otimes I)+(A\otimes I)S}{(n+2)(n+1)n},$$  where $S\in\mathcal{B(H\otimes H)}$ is the swap operator.

{\bf Proof.} $(a)$ It is clear that $$\int_{\mathcal{S}} A\phi\phi^*\otimes B\phi\phi^*\otimes\phi\phi^*d\phi
 =(A\otimes B\otimes I)\int_{\mathcal{S}}\phi\phi^*\otimes\phi\phi^*\otimes\phi\phi^*d\phi
 =\frac{\sum_{i=1}^6(A\otimes B\otimes I)\Gamma(s_i)}{(n+2)(n+1)n},$$ where $\Gamma(s_i)$ is the same as above.
 Then \begin{equation}\int_{\mathcal{S}}\langle A\phi,\phi\rangle\langle B\phi,\phi\rangle\phi\phi^*d\phi
 =\frac{\sum_{i=1}^6{\rm tr}_{12}[(A\otimes B\otimes I)\Gamma(s_i)]}{(n+2)(n+1)n}.\end{equation}
Using a direct calculation, we get that \begin{equation}{\rm tr}_{12}[(A\otimes B\otimes I)\Gamma(s_1)]={\rm tr}(A){\rm tr}(B)I,\end{equation}
 \begin{equation}{\rm tr}_{12}[(A\otimes B\otimes I)\Gamma(s_2)]={\rm tr}(A)B \ \hbox{ and }\ {\rm tr}_{12}[(A\otimes B\otimes I)\Gamma(s_3)]={\rm tr}(AB)I.\end{equation}  Suppose that $\{e_i\}_{i=1}^n$ is an orthonormal basis of $\mathcal{H}.$
   It is obvious that  $$\Gamma(s_4)(x_1\otimes x_2\otimes x_3)=x_3\otimes x_2\otimes x_1=(\sum\limits_{i,j=1}^{n}e_ie_j^*\otimes I\otimes e_je_i^*)(x_1\otimes x_2\otimes x_3)$$ for all $x_1\otimes x_2\otimes x_3\in\mathcal{H}^{\otimes 3},$ so $\Gamma(s_4)=\sum\limits_{i,j=1}^{n}e_ie_j^*\otimes I\otimes e_je_i^*,$ which yields that \begin{equation}\begin{array}{rcl}{\rm tr}_{12}[(A\otimes B\otimes I)\Gamma(s_4)]&=&{\rm tr}_{12}[(A\otimes B\otimes I)(\sum\limits_{i,j=1}^{n}e_ie_j^*\otimes I\otimes e_je_i^*)]\\&=&\sum\limits_{i,j=1}^{n}{\rm tr}(A e_ie_j^*){\rm tr}(B)e_je_i^*={\rm tr}(B)A.\end{array}\end{equation} Moreover, for all $x,y\in\mathcal{H},$ there are $$\begin{array}{rcl}\langle {\rm tr}_{12}[(A\otimes B\otimes I)\Gamma(s_5)]x,y\rangle&=&\sum\limits_{i,j=1}^{n}\langle[(A\otimes B\otimes I)\Gamma(s_5)](e_i\otimes e_j\otimes x),e_i\otimes e_j\otimes y\rangle\\&=&\sum\limits_{i,j=1}^{n}\langle(A\otimes B\otimes I)( x \otimes e_i\otimes e_j),e_i\otimes e_j\otimes y\rangle\\&=&\sum\limits_{i,j=1}^{n}\langle Ax,e_i\rangle\langle Be_i,e_j\rangle \langle e_j,y\rangle=\langle BAx,y\rangle,\end{array}$$ so
   \begin{equation}{\rm tr}_{12}[(A\otimes B\otimes I)\Gamma(s_5)]=BA.\end{equation} In a similar way, we have \begin{equation}{\rm tr}_{12}[(A\otimes B\otimes I)\Gamma(s_6)]=AB.\end{equation}
 Combining equations (3.10)-(3.15), we conclude that $(a)$ holds as desired.

The proof of $(b)$ is similar to that of (a). Indeed, \begin{equation}\int_{\mathcal{S}}\langle A\phi,\phi\rangle\phi\phi^*\otimes\phi\phi^*d\phi
 =\frac{\sum_{i=1}^6{\rm tr}_{1}[(A\otimes I\otimes I)\Gamma(s_i)]}{(n+2)(n+1)n}.\end{equation}
 Moreover,  \begin{equation}{\rm tr}_{1}[(A\otimes I\otimes I)(\Gamma(s_1)+\Gamma(s_2))]={\rm tr}(A)(I\otimes I+S)
  \ \hbox{ and }\ {\rm tr}_{1}[(A\otimes I\otimes I)\Gamma(s_3)]=A\otimes I.\end{equation}
 Also, a direct calculation yields that \begin{equation}{\rm tr}_{1}[(A\otimes I\otimes I)\Gamma(s_4)]=I\otimes A,
\  \  \ {\rm tr}_{1}[(A\otimes I\otimes I)\Gamma(s_5)]=(A\otimes I)S\end{equation}
and $$ {\rm tr}_{1}[(A\otimes I\otimes I)\Gamma(s_6)]=S(A\otimes I).$$   $\Box$

{\bf Remark 3.} Let $S\in\mathcal{B(H\otimes H)}$ be the swap operator and $A,B\in\mathcal{B(H)}.$ A similar proof as above yields that $$\int_{\mathcal{S}}\langle A\phi,\phi\rangle\langle B\phi,\phi\rangle\phi\phi^*\otimes\phi\phi^*d\phi
 =\frac{\sum\limits_{s\in\mathbb{S}_4}{\rm tr}_{12}[(A\otimes B\otimes I\otimes I)\Gamma(s)]}{(n+3)(n+2)(n+1)n}.$$ Here $${\rm tr}_{12}(T)=\sum\limits_{j=1}^{m}{\rm tr}(A_j){\rm tr}(B_j) C_j \otimes D_j$$ for $$T=\sum_{j=1}^{m}A_j\otimes B_j \otimes C_j\otimes D_j\in\mathcal{B(H}^{\otimes 4}).$$
Using a direct calculation, we get that $$\begin{array}{rcl}\int_{\mathcal{S}}\langle A\phi,\phi\rangle\langle B\phi,\phi\rangle\phi\phi^*\otimes\phi\phi^*d\phi
 &=&\frac{(I\otimes I+S)[({\rm tr}(A){\rm tr}(B)+{\rm tr}(AB)+{\rm tr}(A)(B\otimes I+I\otimes B)+{\rm tr}(B)(A\otimes I+I\otimes A)]}{(n+3)(n+2)(n+1)n}\\&+&\frac{(I\otimes I+S)(AB\otimes I+I\otimes AB+BA\otimes I+I\otimes BA+A\otimes B+B\otimes A)}{(n+3)(n+2)(n+1)n}.\end{array}$$

\section{Proofs of Theorem 2 and Theorem 1}

{\bf Proof of Theorem 2.}  Sufficiency is obvious. Necessity. Since $\Phi:\mathcal{T(H)}\longrightarrow\mathcal{T(K)}$ is a completely positive map, it follows from a generalized Choi theorem ([9,15]) that $\Phi(X)=\sum\limits_{i=1}^{\infty}A_iXA_i^*$ and $\sum\limits_{i=1}^{\infty}A_i^*A_i\leq MI$ in the strong operator topology, where $0<M<\infty$ and $A_i\in\mathcal{B(H,K)}$ for $i=1,2,\cdots.$ It follows from Russo-Dye theorem in [22] that
$$\|\Phi\|_{1\rightarrow 1}=\|\sum\limits_{i=1}^{\infty}A_i^*A_i\|\leq M,$$  where $$\|\Phi\|_{1\rightarrow 1}=\sup\{\|\Phi(X)\|_{1}: X\in\mathcal{T(H)} \hbox{ with } \|X\|_{1}\leq1\}.$$ For any positive operator $A\in\mathcal{T(H)},$ we conclude from the spectral decomposition theorem  that
$A=\sum_{j=1}^{\infty}\lambda_j x_j x_j^*,$ where $\lambda_j\geq\lambda_{j+1}\geq0$ and $\{x_j\}_{j=1}^\infty$ is an orthonormal basis of ${\mathcal{H}}.$ Then $$\begin{array}{rcl}\lim\limits_{n\rightarrow\infty}|{\rm tr}[\Phi(A)-\sum_{j=1}^{n}\lambda_j]|
&=&\lim\limits_{n\rightarrow\infty}|{\rm tr}[\Phi(A-\sum_{j=1}^{n}\lambda_j x_j x_j^*)]|\\&\leq&\lim\limits_{n\rightarrow\infty}\|\Phi(A-\sum_{j=1}^{n}\lambda_j x_j x_j^*)\|_1\\&\leq& M\lim\limits_{n\rightarrow\infty}\|A-\sum_{j=1}^{n}\lambda_j x_j x_j^*\|_1=0,\end{array}$$   so $${\rm tr}(A)=\sum_{j=1}^{\infty}\lambda_j={\rm tr}[\Phi(A)].$$ By using the linearity of $\Phi,$ we get that $${\rm tr}[\Phi(X)]={\rm tr}(X) \ \ \ \hbox{ for all } \ X\in\mathcal{T(H)},$$ so $\sum\limits_{i=1}^{\infty}A_i^*A_i=I.$
 In the following, we denote by $ker(A)$ and $\overline{ran(A)}$ the null space and the closure of range of an operator $A,$ respectively.

{\bf Case 1.} If all $rank(A_i)=1,$ then $A_i=y_iz_i^*,$ where $y_i,z_i\in\mathcal{H}\backslash\{0\}$ for $i=1,2,\cdots.$ Since $\Phi(xx^*)$ are pure states for all unit vectors $x\in\mathcal{H,}$ it follows that there exist a unit vector $\psi$ and complex numbers $\mu_i$ such that $y_i=\mu_i\psi$ for $i=1,2,\cdots.$ Thus
$$I=\sum\limits_{i=1}^{\infty}A_i^*A_i=\sum\limits_{i=1}^{\infty}(y_iz_i^*)^*y_iz_i^*=
\sum\limits_{i=1}^{\infty}y_i^*y_i(z_iz_i^*)=\sum\limits_{i=1}^{\infty}|\mu_i|^2z_iz_i^*.$$ So for all $X\in\mathcal{T(H)},$ there are $${\rm tr}(X)={\rm tr}[\sum\limits_{i=1}^{\infty}|\mu_i|^2z_iz_i^*X]=\sum\limits_{i=1}^{\infty}|\mu_i|^2\langle Xz_i, z_i\rangle$$ and
$$\Phi(X)=\sum\limits_{i=1}^{\infty}A_iXA_i^*=\sum\limits_{i=1}^{\infty}|\mu_i|^2\langle Xz_i, z_i\rangle \psi\psi^*={\rm tr}(X)\psi\psi^*.$$

{\bf Case 2.} If at least one $rank(A_i)>1,$ then without loss of generality, we assume $rank(A_1)>1.$ For any
$0\neq x\in ker(A_1)^\perp,$ we get that there exist complex numbers  $t_{ix}\in\mathbb{C}$ such that $A_ix=t_{ix}A_1x,$ because $\frac{1}{\|x\|^2}\sum\limits_{i=1}^{\infty}A_ix(A_ix)^*$ is a pure state. It follows from linearity of operators $A_i$ that there are complex numbers  $t_{i}\in\mathbb{C}$ such that $A_ix=t_{i}A_1x$  for all $ x\in ker(A_1)^\perp$ and $i=2,3,\cdots.$ Hence, with respect to the space decomposition $\mathcal{H}=ker(A_1)^\perp\oplus ker(A_1)$ and $\mathcal{K}=\overline{ran(A_1)}\oplus ran(A_1)^\perp,$
$$A_1=\left(\begin{array}{cc}A_{11}&0\\ 0&0\end{array}\right)\ \ \hbox{ and } \ \ \ A_i=\left(\begin{array}{cc}t_iA_{11}&A_{i2}\\ 0&A_{i3}\end{array}\right).$$
Let $x_1\in ker(A_1)^\perp$ be a fixed unit vector and $x_2\in ker(A_1)$ be any unit vector.
 Then  $$\Phi[(\frac{x_1}{\sqrt{2}}+\frac{x_2}{\sqrt{2}})(\frac{x_1}{\sqrt{2}}+\frac{x_2}{\sqrt{2}})^*]
 =\sum\limits_{i=1}^{\infty}A_i(\frac{x_1}{\sqrt{2}}+\frac{x_2}
 {\sqrt{2}})[A_i(\frac{x_1}{\sqrt{2}}+\frac{x_2}{\sqrt{2}})]^*$$ is a pure state, which implies that there exist complex numbers  $\mu_i\in\mathbb{C}$ for $i=2,3,\cdots,$ such that
 $$A_i(\frac{x_1}{\sqrt{2}}+\frac{x_2}{\sqrt{2}})=\mu_iA_1(\frac{x_1}{\sqrt{2}}+
 \frac{x_2}{\sqrt{2}})=\frac{\mu_i A_{11} x_1}{\sqrt{2}},$$ and hence
 $$A_{i3}x_2=(\mu_i-t_i) A_{11} x_1-A_{i2}x_2\in\overline{ ran(A_1)}\cap ran(A_1)^\perp.$$ Thus $A_{i3}x_2=0,$ so $A_{i3}=0$ follows from the fact that
$x_2\in ker(A_1)$ is arbitrary.

Since $rank(A_1)>1,$ it follows that there exists $x_3\in ker(A_1)^\perp$ such that $A_{11}x_1$ and $A_{11}x_3$ are linear independent. As shown above, we can get that there exist complex numbers $\nu_i\in\mathbb{C}$ such that
$$(\mu_i-t_i) A_{11} x_1=A_{i2}x_2=(\nu_i-t_i) A_{11} x_3.$$ Thus $A_{i2}x_2=0,$ which yields that
 $A_{i2}=0$ for $i=2,3,\cdots,$ so $A_i=\left(\begin{array}{cc}t_iA_{11}&0\\ 0&0\end{array}\right).$
 Then $\sum\limits_{i=1}^{\infty}A_i^*A_i=I$ implies that $$ ker(A_1)=\{0\}\  \ \hbox{ and }\ \ (\sum\limits_{i=2}^{\infty}|t_i|^2+1)A_1^*A_1=I.$$ Defining $V=\sqrt{\sum\limits_{i=2}^{\infty}|t_i|^2+1}A_1,$
we conclude that $V$ is an isometry and $$\Phi(X)=\sum\limits_{i=1}^{\infty}A_iXA_i^*=(\sum\limits_{i=2}^{\infty}|t_i|^2+1)A_1 XA_1^*=VXV^*.$$  $\Box$

{\bf Remark 4.} Let $\mathcal{H}$ and $\mathcal{K}$ be finite dimensional Hilbert spaces with $\dim\mathcal{K}<\dim\mathcal{H}.$ If $\Phi:\mathcal{B(H)}\longrightarrow\mathcal{B(K)}$ is a completely positive map, then Theorem 2 implies that $\Phi(xx^*)$ are pure states for all unit vectors $x\in\mathcal{H}$ if and only if  $\Phi(\cdot)={\rm tr}(\cdot)\psi\psi^*$ is a pure-state replacement channel, where $\psi\in\mathcal{K}$ is a unit vector.

{\bf Proof of Theorem 1.} We conclude from Choi theorem ([4]) that $\mathcal{E}(X)=\sum\limits_{i=1}^{m}A_iXA_i^*$ with $m\leq dn.$  Let $\{X_j : j=1, 2,\cdots, n^2\}$ be an orthonormal basis of $\mathcal{B(H)}.$
It is easy to verify that  $$\sum\limits_{j=1}^{n^2}{\rm tr}(AX_j^*BX_j)={\rm tr}(A){\rm tr}(B)$$ for all $A,B\in\mathcal{B(H)}$ ([17, Lemma 1]). Thus \begin{equation}\begin{array}{rcl}\|\mathcal{E}\|_2^2=\sum\limits_{j=1}^{n^2}\|\mathcal{E}(X_j)\|_2^2&=&\sum\limits_{j=1}^{n^2}
{\rm tr}[\mathcal{E}^*\mathcal{E}(X_j^*)X_j]\\&=&\sum\limits_{i=1}^{m}\sum\limits_{l=1}^{m}
\sum\limits_{j=1}^{n^2}{\rm tr}(A_i^* A_lX_j^*A_l^*A_iX_j)\\&=&\sum\limits_{i=1}^{m}\sum\limits_{l=1}^{m}
{\rm tr}(A_i^*A_l){\rm tr}(A_l^*A_i)\\&=&\sum\limits_{i=1}^{m}\sum\limits_{l=1}^{m}
|{\rm tr}(A_i^*A_l)|^2\end{array}\end{equation} and
\begin{equation} \begin{array}{rcl}\|\widetilde{\mathcal{E}}\|_2^2=\sum\limits_{k=1}^{n^2}\|\widetilde{\mathcal{E}}(X_k)\|_2^2
 &=&\sum\limits_{k=1}^{n^2}\sum\limits_{i,j=1}^{m}|{\rm tr}(A_iX_kA_j^*)|^2
 \\&=&\sum\limits_{i,j=1}^{m}\sum\limits_{k=1}^{n^2}|{\rm tr}(A_j^*A_iX_k)|^2
 \\&=&\sum\limits_{i,j=1}^{m}{\rm tr}(A_j^*A_iA_i^*A_j)\\&=&{\rm tr}(\mathcal{E}^*\mathcal{E}(I))={\rm tr}(\mathcal{E}(I)^2).
 \end{array}\end{equation}

$(1)$ Using Proposition 3 $(d)$, we get that  $$\begin{array}{rcl}\int_{\mathcal{S}}[\mathcal{E}(\phi\phi^*)]^2d\phi&=&
\int_{\mathcal{S}}[\sum\limits_{i=1}^{m}A_i(\phi\phi^*)A_i^*]^2d\phi\\&=&
\int_{\mathcal{S}}\sum\limits_{i,j=1}^{m}A_i(\phi\phi^*)A_i^*A_j(\phi\phi^*)A_j^*d\phi\\&=&
\sum\limits_{i,j=1}^{m}A_i(\int_{\mathcal{S}}\langle A_i^*A_j\phi,\phi\rangle\phi\phi^*d\phi) A_j^*\\&=&
\sum\limits_{i,j=1}^{m}A_i( \frac{A_i^*A_j+{\rm tr}(A_i^*A_j)I}{n(n+1)})A_j^*\\&=&
\sum\limits_{i,j=1}^{m} \frac{A_iA_i^*A_j A_j^*+{\rm tr}(A_i^*A_j)A_iA_j^*}{n(n+1)}.\end{array}$$ Then \begin{equation}\int_{\mathcal{S}}{\rm tr}[\mathcal{E}(\phi\phi^*)^2]d\phi={\rm tr}[\int_{\mathcal{S}}(\mathcal{E}(\phi\phi^*))^2d\phi]
=\frac{\|\mathcal{E}\|_2^2+\|\widetilde{\mathcal{E}}\|_2^2}{n(n+1)}.\end{equation}
Since $\mathcal{E}(\phi\phi^*)\in\mathcal{B(K)}$ is a positive operator with ${\rm tr}[\mathcal{E}(\phi\phi^*)]=1$ and  $\dim\mathcal{K}=d,$ it follows that $$\frac{1}{d}\leq {\rm tr}[\mathcal{E}(\phi\phi^*)^2]\leq 1$$ for all $\phi\in\mathcal{S},$ so equation (4.3) implies $$\frac{1}{d}\leq\frac{\|\mathcal{E}\|_2^2+\|\widetilde{\mathcal{E}}\|_2^2}{n(n+1)}\leq 1.$$
Thus  $$\frac{n+n^2}{d}\leq\|\mathcal{E}\|_2^2+\|\widetilde{\mathcal{E}}\|_2^2\leq n^2+n.$$

$(2)$ $(a)\Longleftrightarrow(b).$  Since $\frac{1}{d}\leq {\rm tr}[\mathcal{E}(\phi\phi^*)^2]$ and ${\rm tr}[\mathcal{E}(\phi\phi^*)^2]$ is a continuous function on the unit sphere $\mathcal{S},$ it follows that $\frac{n+n^2}{d}=\|\mathcal{E}\|_2^2+\|\widetilde{\mathcal{E}}\|_2^2$ if and only if ${\rm tr}[\mathcal{E}(\phi\phi^*)^2]=\frac{1}{d}$ for all $\phi\in\mathcal{S},$ which is equivalent to
$\mathcal{E}(\phi\phi^*)=\frac{1}{d}I$ for all $\phi\in\mathcal{S}.$ By imposing the linearity condition of $\mathcal{E},$ we conclude that $\mathcal{E}(\phi\phi^*)=\frac{1}{d}I$ for all $\phi\in\mathcal{S}$ if and only if $\mathcal{E}(X)={\rm tr}(X)\frac{I}{d}$ for all $X\in \mathcal{B(H)}.$

$(b)\Longrightarrow(c).$ If $\mathcal{E}(X)={\rm tr}(X)\frac{I}{d},$ then we get from equations (4.1)-(4.2) that $\|\mathcal{E}\|_2=\sqrt{\frac{n}{d}}$ and $\|\widetilde{\mathcal{E}}\|_2=\frac{n}{\sqrt{d}},$ so
$\|\mathcal{E}\|_2+\|\widetilde{\mathcal{E}}\|_2=\frac{\sqrt{n}+n}{\sqrt{d}}.$ 

$(c)\Longrightarrow(d).$ When $\|\mathcal{E}\|_2+\|\widetilde{\mathcal{E}}\|_2=\frac{\sqrt{n}+n}{\sqrt{d}}$ holds, we conclude from inequalities (1.7) and (1.8) that $\|\mathcal{E}\|_2^2=\frac{n}{ d }$ and $\|\widetilde{\mathcal{E}}\|_2^2=\frac{n^2}{d},$  so $\|\mathcal{E}\|_2=\sqrt{\frac{n}{ d }}.$

$(d)\Longrightarrow(b).$ If $\|\mathcal{E}\|_2=\sqrt{\frac{n}{ d }},$  then equation (4.1) implies  $$\sum\limits_{i=1}^{m}\sum\limits_{l=1}^{m}
|{\rm tr}(A_i^*A_l)|^2=\|\mathcal{E}\|_2^2= \frac{n}{ d }.$$ Since the matrix $[{\rm tr}(A_i^*A_j)]_{i,j=1}^{m}$ is positive, it follows that there exists a unitary matrix $U=[\mu_{ij}]_{i,j=1}^{m}$ such that \begin{equation}U^*[{\rm tr}(A_i^*A_j)]_{i,j=1}^{m}U=dig(\lambda_1,\lambda_2,\cdots,\lambda_m).\end{equation} Setting $E_i=\sum_{j=1}^{m}\mu_{ji}A_j$ for $i=1,2,\cdots,m,$ we get from equation (4.4) that $$\mathcal{E}(X)=\sum\limits_{i=1}^{m}A_iXA_i^* =\sum\limits_{i=1}^{m}E_iXE_i^*  \ \  \hbox{ and }\ \ {\rm tr}(E_i^*E_j)=0 \ \hbox{ for } i\neq j.$$  Thus $$\sum\limits_{i=1}^{m}
[\frac{{\rm tr}(E_i^*E_i)}{n}]^2=\frac{\|\mathcal{E}\|_2^2}{n^2}=\frac{1}{ d n }\ \ \hbox{ and }\ \ \sum\limits_{i=1}^{m}
\frac{{\rm tr}(E_i^*E_i)}{n}=\frac{1}{n}{\rm tr}(\sum\limits_{i=1}^{m}
E_i^*E_i)=1.$$ Then inequality  $m\leq dn$ and  Cauchy-Schwarz inequality imply that $$1=(\sum\limits_{i=1}^{m}
\frac{{\rm tr}(E_i^*E_i)}{n})^2\leq m\sum\limits_{i=1}^{m}
[\frac{{\rm tr}(E_i^*E_i)}{n}]^2=\frac{m}{dn}\leq1,$$ so $m=dn$ and ${\rm tr}(E_i^*E_i)=\frac{1}{d}$ for $i=1,2,\cdots,dn.$
Let $F_i=\sqrt{d}E_i$ for $i=1,2,\cdots,dn.$ Clearly, $\{F_j : j=1, 2,\cdots, dn\}$ is an orthonormal basis of $\mathcal{B(H,K)}.$ Hence [17, Lemma 1] yields  $$\mathcal{E}(X)=\sum\limits_{i=1}^{dn}E_iXE_i^*=\frac{1}{d}\sum\limits_{i=1}^{dn}F_iXF_i^*={\rm tr}(X)\frac{I}{d}.$$

$(3)$ In a similar way to the above $(2),$ ${\rm tr}[\mathcal{E}(\phi\phi^*)^2]\leq 1$ and the continuity  of ${\rm tr}[\mathcal{E}(\phi\phi^*)^2]$ on the the unit sphere $\mathcal{S}$ imply that $n+n^2=\|\mathcal{E}\|_2^2+\|\widetilde{\mathcal{E}}\|_2^2$ if and only if ${\rm tr}[\mathcal{E}(\phi\phi^*)^2]=1$ for all $\phi\in\mathcal{S},$ which is equivalent to that
$\mathcal{E}(\phi\phi^*)$ are pure states for all $\phi\in\mathcal{S}.$  Then Theorem 2 yields that $\|\mathcal{E}\|_2^2+\|\widetilde{\mathcal{E}}\|_2^2=n^2+n$ if and only if
$\mathcal{E}(X)=VXV^*$  or  $\mathcal{E}(X)={\rm tr}(X)\psi\psi^*,$ where $V\in\mathcal{B(H,K)}$ is an isometry and $\psi\in\mathcal{K}$ is a unit vector.  $\Box$

{\bf Remark 5.} Let $\Gamma$ denote the set of all quantum channels from $\mathcal{B(H)}$ into $\mathcal{B(K)}.$  Then we claim that the range of all possible values of $\|\mathcal{E}\|_2^2+\|\widetilde{\mathcal{E}}\|_2^2$ for all quantum channels $\mathcal{E}\in\Gamma$ is the closed interval $[\frac{n+n^2}{d}, \ n^2+n],$ where $\dim\mathcal{H}=n$ and $\dim\mathcal{K}=d.$  That is,  \begin{equation}\{\|\mathcal{E}\|_2^2+\|\widetilde{\mathcal{E}}\|_2^2:\ \
\mathcal{E} \in \Gamma\}=[\frac{n+n^2}{d}, \ n^2+n].\end{equation} Indeed, define quantum channels $\mathcal{E}_\lambda$ on the closed interval $[0,1]$ by
$$\mathcal{E}_\lambda(X)=\lambda {\rm tr}(X)\psi\psi^*+(1-\lambda){\rm tr}(X)\frac{I}{d},$$ where $\psi\in\mathcal{K}$ is a fixed unit vector. Let $\{e_j: j=1, 2,\cdots, n\}$ be an orthonormal basis of $\mathcal{H}.$ Then
 $$\|\mathcal{E}_\lambda\|_2^2=\sum\limits_{i=1}^{n}\sum\limits_{j=1}^{n}\|\mathcal{E}_\lambda(e_ie_j^*)\|_2^2
 =\sum\limits_{i=1}^{n}\|\lambda  \psi\psi^*+(1-\lambda)\frac{I}{d}\|_2^2=\frac{n(1-\lambda^2)}{d}+n\lambda^2 $$ and $$\|\widetilde{\mathcal{E}_\lambda}\|_2^2={\rm tr}[\mathcal{E}_\lambda(I)^2]=\frac{n^2(1-\lambda^2)}{d}+n^2\lambda^2.$$ Clearly,
 $$\|\mathcal{E}_0\|_2^2+\|\widetilde{\mathcal{E}_0}\|_2^2=\frac{n+n^2}{d}  \ \hbox{ and } \ \
 \|\mathcal{E}_1\|_2^2+\|\widetilde{\mathcal{E}_1}\|_2^2=n+n^2.$$
 Since $\|\mathcal{E}_\lambda\|_2^2+\|\widetilde{\mathcal{E}_\lambda}\|_2^2$ is a continuous function on the closed interval $[0,1],$ it follows that $$\{\|\mathcal{E}_\lambda\|_2^2+\|\widetilde{\mathcal{E}_\lambda}\|_2^2: \ \lambda\in[0,1]\}\supseteq[\frac{n+n^2}{d},\ n^2+n].$$ So Theorem 1 implies the desired equation (4.5). $\Box$

Recall that an isometric channel specifically refers to maps of the form $\rho\rightarrow V\rho V^\dag,$
where $V\in B\mathcal{(H,K)}$ is an isometry ([28]). In this case, $\dim\mathcal{H}\leq \dim\mathcal{K}.$
A channel $\mathcal{E}:
B\mathcal{(H})\rightarrow B\mathcal{ (K})$ is called a random isometric channel, if  $\mathcal{E}(X)=\sum\limits_{j=1}^{m}p_jV_jX V_j^\dag,$ where $V_j\in B\mathcal{(H,K})$ are isometries and $\{p_j\}_{j=1}^m$ is a probability distribution. Particularly, $\mathcal{E}:
B\mathcal{(H})\rightarrow B\mathcal{ (K})$ is called a random unitary channel, if $\dim\mathcal{H}= \dim\mathcal{K} $ and these $V_j\in B\mathcal{(H,K})$ are unitary operators ([28]). In the following, we consider the range of all possible values of $\|\mathcal{E}\|_2^2+\|\widetilde{\mathcal{E}}\|_2^2$ for all random isometric channels $\mathcal{E}.$

{\bf Corollary 10.} Let $\dim\mathcal{H}=n\leq d=\dim\mathcal{K}$ and $\Delta$ be the set of all random isometric channels from $\mathcal{B(H)}$ into $\mathcal{B(K)}.$

$(a)$ If $\mathcal{E}:\mathcal{B(H)}\longrightarrow\mathcal{B(K)}$ is a random isometric channel and $\widetilde{\mathcal{E}}$ is a complementary channel of $\mathcal{E},$ then $\frac{n^2}{d}\leq\|\widetilde{\mathcal{E}}\|_2^2\leq n,$ and $\|\widetilde{\mathcal{E}}\|_2^2= n$ if and only if there exist an isometry $V\in\mathcal{B(H,K)}$ and a random unitary channel $\Upsilon: \mathcal{B(H)}\longrightarrow\mathcal{B(H)}$ such that $\mathcal{E}(X)=V\Upsilon(X)V^*.$

$(b)$  The completely depolarizing channel $\mathcal{E}_c:\mathcal{B(H)}\longrightarrow\mathcal{B(K)}$ with $\mathcal{E}_c(X)=\frac{{\rm tr}(X)I}{d}$ is a random isometric channel.

$(c)$   $\{\|\mathcal{E}\|_2^2+\|\widetilde{\mathcal{E}}\|_2^2: \ \
\mathcal{E}\in\Delta\}=[\frac{n+n^2}{d}, \ n^2+n].$

{\bf Proof.} $(a)$  Using equation (4.2), we get that \begin{equation}\|\widetilde{\mathcal{E}}\|_2^2= {\rm tr}[\mathcal{E}(I)^2]\geq\frac{{\rm tr}[\mathcal{E}(I)]^2}{d}=\frac{n^2}{d}.\end{equation} 
Suppose that $\mathcal{E}(X)=\sum\limits_{j=1}^{m}p_jV_jX V_j^*,$ where all $V_j\in\mathcal{B(H,K})$ are isometries and all $p_j>0$ with $\sum_{j=1}^m p_j=1.$   Then equation (4.2) implies
\begin{equation}\begin{array}{rcl}\|\widetilde{\mathcal{E}}\|_2^2=
\sum\limits_{j=1}^{m}\sum\limits_{i=1}^{m}p_jp_i{\rm tr}(  V_i^*V_jV_j^*V_i )&=&
\sum\limits_{j=1}^{m}\sum\limits_{i=1}^{m}p_jp_i\| V_j^*V_i\|_2^2 \\&\leq&
\sum\limits_{j=1}^{m}\sum\limits_{i=1}^{m}p_jp_i\| V_j^*\|^2\| V_i\|_2^2\\&=&
\sum\limits_{j=1}^{m}\sum\limits_{i=1}^{m}p_jp_in=n.\end{array}\end{equation}
It follows from inequality (4.7) that $\|\widetilde{\mathcal{E}}\|_2^2= n$ yields
$$\sum\limits_{j=1}^{m}\sum\limits_{i=1}^{m}p_jp_i\| V_j^* V_i\|_2^2 =
\sum\limits_{j=1}^{m}\sum\limits_{i=1}^{m}p_jp_i\| V_j^*\|^2\| V_i\|_2^2,$$ so
$$\| V_j^* V_i\|_2^2=\| V_j^*\|^2\| V_i\|_2^2=n\ \ \hbox{ for }\ i,j=1,2,\cdots,m.$$ Thus
$${\rm tr}(V_i V_i^*V_jV_j^*)={\rm tr}(V_i^*V_j V_j^* V_i)=n={\rm tr}(V_jV_j^*)={\rm tr}(V_iV_i^*),$$
and hence, $$V_jV_j^* =V_i V_i^*\ \ \hbox{ and }\ \ V_i^* V_j V_j^*V_i =I.$$
Then all $V_i^*V_j\in\mathcal{B(H})$ are unitary operators and  $$V_j=V_jV_j^*V_j=V_iV_i^* V_j.$$  Setting $U_1=I$ and $U_j=V_1^* V_j$ for $j=2,3,\cdots, m,$ we get that
 $V_j=V_1U_j$ and $$\mathcal{E}(X)=\sum\limits_{j=1}^{m}p_jV_jX V_j^*=p_1V_1XV_1^*+\sum\limits_{j=2}^{m}p_jV_1U_jXU_j^* V_1^*=V\Upsilon(X)V^*,$$  where $V=V_1\in\mathcal{B(H,K)}$ is an isometry and $\Upsilon(X)=\sum\limits_{j=1}^{m}p_jU_jXU_j^*$ is
 a random unitary channel.

 For the  other direction,  we conclude from equation (4.2) that $$\|\widetilde{\mathcal{E}}\|_2^2={\rm tr}[\mathcal{E}(I)^2]={\rm tr}[(p_1
V_1V_1^*+\sum\limits_{j=2}^{m}p_jV_1U_jU_j^*V_1^*)^2]={\rm tr}(
V_1V_1^*)=n.$$

$(b)$ Let $\{e_j: j=1, 2,\cdots, n\}$ and $\{f_j: j=1, 2,\cdots, d\}$ be orthonormal bases of $\mathcal{H}$ and $\mathcal{K}$ respectively.  We define an isometry $W\in\mathcal{B(H,K})$ and two unitary operators $U\in \mathcal{B(K})$ and
 $V\in\mathcal{B(H})$ as
$$W=\sum\limits_{k=1}^{n}f_ke_k^*,\  \  \ U=\sum\limits_{i=1}^{d-1}f_{i+1}f_i^*+f_1f_{d}^* \ \ \ \hbox{ and }\ \ V=\sum\limits_{j=1}^{n} \omega^{j-1}e_je_j^*,$$ where $\omega=\mathrm{e}^{\frac{2\pi\sqrt{-1}}{n}}$ is the principal $n$-th root of unity. Denote by $V_{ij}=U^jWV^i$ for $i=1,2,\cdots,n$ and $j=1,2,\cdots,d.$
 Clearly, $W^*W=I,$ and hence  $$V_{ij}^* V_{ij}=(V^i)^* W^*(U^j)^* U^jWV^i=I.$$ Moreover, if $i\neq k=1,2,\cdots,n$ or $j\neq l=1,2,\cdots,d,$ then $$ \langle V_{kl}, V_{ij}\rangle={\rm tr}[(V^i)^* W^*(U^j)^* U^lWV^k]={\rm tr}[W^* U^{l-j}WV^{k-i}]={\rm tr}[U^{l-j}WV^{k-i}W^*]=0.$$ So all $V_{ij}$ are isometries and
 $\{\frac{ V_{ij}}{\sqrt{n}}: i=1,2,\cdots,n\ \hbox{ and } \ j=1, 2,\cdots,d\}$ is an orthonormal basis of $\mathcal{B(H,K})$ with respect to the Hilbert-Schmidt inner product. Thus we get from [17, Lemma 1] that the completely depolarizing channel $$\mathcal{E}_c(X)=\frac{{\rm tr}(X)I}{d}=\frac{1}{nd}\sum\limits_{i=1}^n\sum\limits_{j=1}^d V_{ij}XV_{ij}^*$$ is a random isometric channel.

$(c)$ Define random isometric channels $\mathcal{E}_t$ on the closed interval $[0,1]$ by
\begin{equation}\mathcal{E}_t(X)=tV_{11}XV_{11}^*+(1-t){\rm tr}(X)\frac{I}{d}=tV_{11}XV_{11}^*+(1-t)\frac{1}{nd}
\sum\limits_{i=1}^n\sum\limits_{j=1}^d V_{ij}XV_{ij}^*,\end{equation} where $V_{ij}\in\mathcal{B(H,K})$ is the same as above $(b).$
We conclude from equations (4.1) and (4.8) that  $$\begin{array}{rcl}\|\mathcal{E}_t\|_2^2&=&(t+\frac{1-t}{nd})^2{\rm tr}(V_{11}V_{11}^*)+\frac{(1-t)^2}{n^2d^2}\{\sum\limits_{i=2}^{n}\sum\limits_{j=1}^{d}
|{\rm tr}(V_{ij}V_{ij}^*)|^2+\sum\limits_{j=2}^{d}
|{\rm tr}(V_{1j}V_{1j}^*)|^2\}
 \\&=&n^2 t^2+ \frac{n(1-t^2)}{d}\end{array}$$ and $$\|\widetilde{\mathcal{E}_t}\|_2^2={\rm tr}[\mathcal{E}_t(I)^2]=\frac{n^2(1-t^2)}{d}+nt^2.$$
 Obviously, $\|\mathcal{E}_t\|_2^2+\|\widetilde{\mathcal{E}_t}\|_2^2$ is a continuous function on the closed interval $[0,1],$  $$\|\mathcal{E}_0\|_2^2+\|\widetilde{\mathcal{E}_0}\|_2^2=\frac{n+n^2}{d}  \  \hbox{ and } \ \ \|\mathcal{E}_1\|_2^2+\|\widetilde{\mathcal{E}_1}\|_2^2=n+n^2.$$
  Then $$\{\|\mathcal{E}\|_2^2+\|\widetilde{\mathcal{E}}\|_2^2:
\mathcal{E}\in\Delta\}\supseteq\{\|\mathcal{E}_t\|_2^2+\|\widetilde{\mathcal{E}_t}\|_2^2:
 t\in[0,1]\}=[\frac{n+n^2}{d},\ n^2+n],$$ so $(c)$ follows from Theorem 1.  $\Box$

\section{An application of Theorem 9}

Recall that the Hermitian-preserving and trace-preserving map
$\mathcal{M}: \mathcal{B(H)}\longrightarrow\mathcal{B(H\otimes H)}$ with
$$\mathcal{M}(X):=n\int_{\mathcal{S}}{\rm tr}(\rho_\phi X)(\rho_\phi \otimes\rho_\phi)d\phi \ \ \ \  \hbox{ for all  } X\in\mathcal{B(H)},$$ where
$\rho_\phi:=\frac{1}{2}[(n+2)\phi\phi^*-I]$ for a unit vector $\phi\in\mathcal{H}$ and $\dim\mathcal{H}=n$ in [19]. Indeed, $${\rm tr}[\mathcal{M}(X)]=n\int_{\mathcal{S}}{\rm tr}[\rho_\phi X]tr[(\rho_\phi \otimes\rho_\phi)]d\phi=n[\frac{n+2}{2}\int_{\mathcal{S}}{\rm tr}(X\phi\phi^*)d\phi-\frac{{\rm tr}(X)}{2}]={\rm tr}(X).$$
Another Hermitian-preserving and trace-preserving map $\mathcal{CB}:\mathcal{B(H)}\longrightarrow\mathcal{B(H\otimes H)}$ with  $$\mathcal{CB}(X):=\frac{S(X\otimes I)+(X\otimes I)S}{2}$$ is called the canonical broadcasting map in [8,19], where $S\in\mathcal{B(H\otimes H)}$ is the swap operator. It is worth noting that the canonical broadcasting map $\mathcal{CB}$ has several nice properties and applications ([19, Theorem 1-4]).

Let $\mathcal{M'}: \mathcal{B(H)}\longrightarrow\mathcal{B(H\otimes H)}$ with $\mathcal{M'}(X):={\rm tr}(X)\frac{I}{n}\otimes \frac{I}{n}$ be the completely depolarizing channel.
Then by using the Jamio{\l}kowski isomorphism ([12]), the following equation is obtained in [19]
 \begin{equation} \mathcal{CB}(X)=p\mathcal{M}(X)+(1-p)\mathcal{M'}(X),\end{equation}  where $p:=\frac{4(n+1)}{(n+2)^2}.$

We shall give a simpler proof of the above equation (5.1). Indeed, by using Proposition 3 and Theorem 9 (b), we get that $$\int_{\mathcal{S}}{\rm tr}(X\phi\phi^*)\phi\phi^*\otimes\phi\phi^*d\phi=\frac{{\rm tr}(X)(I\otimes I+S)+I\otimes X+X\otimes I+S(X\otimes I)+(X\otimes I)S}{(n+2)(n+1)n}$$ and $$\int_{\mathcal{S}}{\rm tr}(X\phi\phi^*)\phi\phi^*d\phi=\frac{X+{\rm tr}(X)I}{n(n+1)}.$$ Then
 a direct calculation yields that $$\begin{array}{rcl}\mathcal{M}(X)&=&n\int_{\mathcal{S}}{\rm tr}(\rho_\phi X)(\rho_\phi \otimes\rho_\phi)d\phi\\&=&\frac{n(n+2)^3}{8}\int_{\mathcal{S}}{\rm tr}(X\phi\phi^*)\phi\phi^*\otimes\phi\phi^*d\phi
+\frac{n(n+2)}{8}(\int_{\mathcal{S}}{\rm tr}(X\phi\phi^*)d\phi) I\otimes I\\&&-\frac{n(n+2)^2}{8}[\int_{\mathcal{S}}{\rm tr}(X\phi\phi^*)\phi\phi^*d\phi\otimes I+I\otimes \int_{\mathcal{S}}{\rm tr}(X\phi\phi^*)\phi\phi^*d\phi]-\frac{n{\rm tr}(X)}{8}I\otimes I\\&&-\frac{n(n+2)^2{\rm tr}(X)}{8}\int_{\mathcal{S}}\phi\phi^*\otimes\phi\phi^*d\phi
+\frac{n(n+2){\rm tr}(X)}{8}[\int_{\mathcal{S}}(\phi\phi^*\otimes I+I\otimes \phi\phi^*)d\phi]\\&=&\frac{(n+2)^2}{8(n+1)}[S(X\otimes I)+(X\otimes I)S]-\frac{{\rm tr}(X)I\otimes I}{4(n+1)}.\end{array}$$
Thus $$\begin{array}{rcl}\mathcal{CB}(X)=\frac{S(X\otimes I)+(X\otimes I)S}{2}&=&\frac{ 4(n+1)}{(n+2)^2}\mathcal{M}(X)+\frac{n^2{\rm tr}(X)}{(n+2)^2}\frac{I}{n}\otimes \frac{I}{n}\\&=&p\mathcal{M}(X)+(1-p)\mathcal{M'}(X).\end{array}$$

{\bf Conflict of interest}  The authors have no conflicts to disclose.

{\bf Data availability statement}
Data sharing is not applicable to this paper as no new data were created or analyzed in this study.


\bibliographystyle{amsplain}

\end{document}